\documentclass[a4paper]{article}

\usepackage{INTERSPEECH2018}
\usepackage{multirow}
\usepackage{graphics}
\usepackage{soul}
\usepackage{xargs}  
\usepackage{array}% Use more than one optional parameter in a new commands
\usepackage{subcaption}
\usepackage[pdftex,dvipsnames]{xcolor}  % Coloured text etc.
\usepackage{tabularx}
\usepackage[colorinlistoftodos,prependcaption,textsize=tiny]{todonotes}
\newcommandx{\unsure}[2][1=]{\todo[linecolor=red,backgroundcolor=red!25,bordercolor=red,#1]{#2}}
\newcommandx{\change}[2][1=]{\todo[linecolor=blue,backgroundcolor=blue!25,bordercolor=blue,#1]{#2}}
\newcommandx{\info}[2][1=]{\todo[linecolor=OliveGreen,backgroundcolor=OliveGreen!25,bordercolor=OliveGreen,#1]{#2}}
\newcommandx{\improvement}[2][1=]{\todo[linecolor=Plum,backgroundcolor=Plum!25,bordercolor=Plum,#1]{#2}}
\newcommandx{\thiswillnotshow}[2][1=]{\todo[disable,#1]{#2}}
\newcolumntype{P}[1]{>{\centering\arraybackslash}p{#1}}
\newcolumntype{K}[1]{>{\centering\arraybackslash}p{#1}}
\usepackage{amsmath}

\title{Integrating Recurrence Dynamics for Speech Emotion Recognition}
\name{Efthymios Tzinis$^{1,2,\dagger}$, 
      Georgios Paraskevopoulos$^{1,2,\dagger}$,
      Christos Baziotis$^{1}$, 
      Alexandros Potamianos$^{1,2}$}
\address{
  $^1$School of Electrical \& Computer Engineering,
       National Technical University of Athens, Greece \\
  $^2$Behavioral Signal Technologies, Los Angeles, CA, USA\\
  $\dagger$Both authors contributed equally to this work}
  
\email{etzinis@gmail.com, geopar@central.ntua.gr, cbaziotis@mail.ntua.gr, potam@central.ntua.gr}

\begin{document}

\maketitle

\begin{abstract}
We investigate the performance of features that can capture nonlinear recurrence dynamics embedded in the speech signal for the task of Speech Emotion Recognition (SER). Reconstruction of the phase space of each speech frame and the computation of its respective Recurrence Plot (RP) reveals complex structures which can be measured by performing Recurrence Quantification Analysis (RQA). These measures are aggregated by using statistical functionals over segment and utterance periods. We report SER results for the proposed feature set on three databases using different classification methods.  When fusing the proposed features with traditional feature sets, e.g., \cite{schuller2010interspeech}, we show an improvement in unweighted accuracy of up to 5.7\% and 10.7\% on Speaker-Dependent (SD) and Speaker-Independent (SI) SER tasks, respectively, over the baseline \cite{schuller2010interspeech}. Following a segment-based approach we demonstrate state-of-the-art performance on IEMOCAP using a Bidirectional Recurrent Neural Network. 
% By employing an attention-based recurrent neural network and combining the two feature sets on segment-level we achieve comparable to state-of-the art performance on IEMOCAP.  

\end{abstract}
\noindent\textbf{Index Terms}: speech emotion recognition, recurrence quantification analysis, nonlinear dynamics, recurrence plots

\section{Introduction}
Automatic Speech Emotion Recognition (SER) is key for building intelligent human-machine interfaces that can adapt to the affective state of the user, especially in cases like call centers where no other information modality is available \cite{shridialogsystems}. 

Extracting features capable of capturing the emotional state of the speaker is a challenging task for SER. Prosodic, spectral and voice quality Low Level Descriptors (LLDs), extracted from speech frames, have been extensively used for SER \cite{el2011survey}. Proposed SER approaches mainly differ on the aggregation and temporal modeling of the input sequence of LLDs. 
In utterance-based approaches, statistical functionals are applied over all LLD values of the included frames \cite{schuller2010interspeech}. 
%In this context, these utterance
These utterance-level statistical representations have been successfully used for SER using Support Vector Machines (SVMs) \cite{shri2011SVM}, Convolutional Neural Networks (CNNs) \cite{emily2017regionalsaliency} and Deep Belief Networks (DBNs) in a multi-task learning setup \cite{xia2017multi}. 
Moreover, segment-based approaches have showcased that computation of statistical functionals over LLDs in appropriate timescales yields a significant performance improvement for SER systems \cite{schuller2006timing}, \cite{tzinis2017segment}. Specifically, in \cite{tzinis2017segment} statistical representations are extracted from overlapping segments, each one corresponding to a couple of words. The resulting sequence of segments representations is fed as input to a Long Short Time Memory (LSTM) unit for SER classification.

Direct SER approaches are usually based on raw LLDs extracted from emotional utterances. CNNs \cite{fayek2017stateoftheart} and Bidirectional-LSTMs (BLSTMs) \cite{ghosh2016representation} over spectrogram representations reported state-of-the-art performances on Interactive Emotional Dyadic Motion Capture (IEMOCAP) database \cite{IEMOCAP}. LSTMs with attention mechanisms have also been proposed in order to accommodate an active selection of the most emotionally salient frames \cite{shri2016attention}, \cite{mirsamadi2017attention}. To this end, Sparse Auto-Encoders (SAE) for learning salient features from  spectrograms of emotional utterances have also been studied \cite{mao2014autoencoder}.      

Despite the great progress that has been made in SER, the aforementioned LLDs are extracted under the assumption of a linear source-filter model of speech generation. However, vocal fold oscillations and vocal tract fluid dynamics often exhibit highly nonlinear dynamical properties which might not be aptly captured by conventional LLDs  \cite{herzel1993bifurcations}. Nonlinear analysis of a speech signal through the reconstruction of its corresponding Phase Space (PS) lies in embedding the signal in a higher dimensional space where its dynamics are unfolded \cite{pitsikalis2009analysis}. Recurrent patterns of these orbits are indicative attributes of system's behavior and can be analyzed using Recurrence Plots (RPs) \cite{RPsfirst1987}. Recurrence Quantification Analysis (RQA) provides complexity measures for an RP which are capable of identifying a system's transitions between chaotic and order regimes \cite{marwan2007recurrenceAll}. A variety of nonlinear features like: Teager Energy Operator \cite{TEOsun2011}, modulation features from instantaneous amplitude and phase \cite{traliarismaragos} as well as geometrical measures from PS orbits \cite{MeasuresonRPS2015speech} have been reported to yield significant improvement on SER when combined with conventional feature sets. However, RQA analysis has not yet been employed for SER. In \cite{SER_RQA_2016exploring} RQA measures have been shown to be statistically significant for the discrimination of emotions but an actual SER experimental setup is missing.

In this paper, we extract RQA measures from speech-frames and evaluate them for SER. We test the efficacy of the proposed RQA feature set under both utterance and segment-based approaches by calculating statistical functionals over the respective time lengths. SVMs and Logistic Regression (LR) classifiers are used for the utterance-based approach as well as an Attention-BLSTM (A-BLSTM) for the respective segment-based approach. The performance of the proposed RQA feature set, as well as the fusion of the RQA features with conventional feature sets \cite{schuller2010interspeech},  is reported on three databases and compared with state-of-the-art results for Speaker-Dependent (SD), Speaker-Independent (SI) and Leave One Session Out (LOSO) SER experiments. 

%Fusion with a traditional feature sets (the IS10 feature set \cite{schuller2010interspeech}) shows that RQA measures substantially improve the performance of all SER models under Speaker-Dependent (SD), Speaker-Independent (SI) and Leave One Session Out (LOSO) experiments. We propose the utilization of RQA feature set as an addition to conventional LLDs for SER tasks. %As a result, we report comparable to state-of-the-art performance in all emotional databases used in evaluation without any further feature selection scheme.  

\section{Feature Extraction}
\label{sect:FeatureExtraction}

\subsection{Baseline Feature Set (IS10 Set)}
\label{sect:IS10featureset}
We use the IS10 feature set \cite{schuller2010interspeech}, in which $1582$ features are extracted corresponding to statistical functionals applied on various LLDs. The extraction is performed for both segment and utterance based approaches using the openSMILE toolkit \cite{opensmile}. 

\subsection{Proposed Nonlinear Feature Set (RQA Set)}
\label{sect:RQAFeatureSet}
The RQA feature set for a given speech segment or utterance is extracted as described next. First, we break the given speech signal into frames and for each one we reconstruct its PS as shown in Section~\ref{sect: Phase Space Reconstruction}. For each PS orbit, its respective RP is computed as explained in Section~\ref{sect: Recurrence Plot Computation}. In order to quantify the complex structures of the RP, a list of RQA measures (described in Section~\ref{sect: RQA measures}) is extracted; resulting in a $12$-dimensional representation of the input speech frame. Representations for speech-segments and utterances containing multiple frames are obtained by applying a set of $18$ statistical functionals (listed in Section~\ref{sect: RQA stats}) over  $12$-dimensional frame-attributes and their deltas. Thus, a $432$-dimensional feature vector is obtained.

\subsubsection{Phase Space Reconstruction}
\label{sect: Phase Space Reconstruction}
Given a speech frame with $N$ samples $\{s(i)\}_{i=1}^{N}$ we reconstruct its corresponding PS trajectory by computing $m$ time-delayed versions of the original speech frame by multiples of time lag $\tau$ and creating the vectors lying in $\mathbb{R}^m$ as shown next:    
\begin{equation}
  \mathbf{x}(i) = [s(i), s(i+\tau),...,s(i+(m-1)\tau)] 
  \label{eq: Phase Space Reconstruction}
\end{equation}
where $m$ is the embedding dimension of the reconstructed PS and $\tau$ is the time lag. If the embedding theorem holds and the aforementioned parameters are set appropriately, then the orbit defined by the points $\{\mathbf{x}(i)\}_{i=1}^{N}$ would truthfully preserve invariant quantities of the true underlying dynamics which are assumed to be unknown \cite{TakensTheorem}. In accordance with \cite{pitsikalis2009analysis}, parameters $\tau$ and $m$ for each speech frame are estimated individually by using Average Mutual Information (AMI) \cite{AMI} and False Nearest Neighbors (FNN) \cite{FalseNearestNeighbors}, respectively.

\subsubsection{Recurrence Plot}
\label{sect: Recurrence Plot Computation}
Given a PS trajectory $\{\mathbf{x}(i)\}_{i=1}^{N}$ we analyze the recurrence properties of these states by calculating the pairwise distances and thresholding these values in order to compute the corresponding RP \cite{RPsfirst1987}. RPs are binary square matrices and are defined element-wise as shown next:
\begin{equation}
  \mathbf{R}_{i,j}(\epsilon, q) = \Theta(\epsilon - ||\mathbf{x}(i) - \mathbf{x}(j)||_q)
  \label{eq: Recurrence Plot}
\end{equation}
where $\Theta(\cdot)$ is the Heaviside step function, $\epsilon$ is the thresholding value, $||\cdot||_q$ is the norm used to define the distance between trajectory points (for $q=1$, $q=2$ or $q=\infty$ we compute Manhattan, Euclidean or Supremum norm, respectively). Thus, matrix $\mathbf{R}$ consists of ones in areas where the states of the orbit are close and zero elsewhere. The measure of proximity is defined by threshold $\epsilon$ for which multiple selection criteria have been studied \cite{schinkel2008eselection}. We consider three criteria depending on: 1) a fixed ad-hoc threshold value, 2) a fixed Recurrence Rate (RR) as defined in Table \ref{t:RQAMeasures} (e.g., For $RR=0.15$ we set $\epsilon$ according to a fixed probability of the pairwise distances of PS's points $P(||\mathbf{x}(i) - \mathbf{x}(j)||_q<\epsilon) = 0.15, \enskip 1 \leq i,j, \leq N$), and 3) a fixed ratio of the standard deviation $\sigma$ of points $\{\mathbf{x}(i)\}_{i=1}^{N}$, e.g., $\epsilon = 5 \sigma$ \cite{stdEselection}. For fixed values of $\epsilon$ and $q$ we denote as $\mathbf{R}_{i,j}$ the respective entry of the RP matrix for simplicity of notation. \newline
An $L$-length diagonal line (of ones) is defined by: 
\begin{equation}
  (1 - \mathbf{R}_{i-1,j-1}) (1 - \mathbf{R}_{i+L+1,j+L+1}) \prod_{k=1}^{k=L}{\mathbf{R}_{i+k,j+k}} = 1
  \label{eq: Diagonal Line}
\end{equation}
An $L$-length vertical line is described by:  
\begin{equation}
  (1 - \mathbf{R}_{i,j-1}) (1 - \mathbf{R}_{i,j+L+1}) \prod_{k=1}^{k=L}{\mathbf{R}_{i,j+k}} = 1
  \label{eq: Vertical Line}
\end{equation}
An $L$-length white vertical line (of zeros) is defined as:  
\begin{equation}
  \mathbf{R}_{i,j-1} \mathbf{R}_{i,j+L+1} \prod_{k=1}^{k=L}{(1-\mathbf{R}_{i,j+k})} = 1
  \label{eq: White Vertical Line}
\end{equation}
We also denote with $P_d(l)$, $P_v(l)$ and $P_{w}(l)$ the histogram distributions of lengths of diagonal, vertical and white vertical lines, respectively. Hence, the total number of these lines are correspondingly $N_d = \sum_{l\geq d_m}P_d(l)$, $N_v = \sum_{l\geq v_m}P_v(l)$ and $N_w = \sum_{l\geq w_m}P_w(l)$, where $d_{m} = 2$, $v_{m} = 2$ and $w_{m} = 1$ define the minimum lengths for each type of line \cite{marwan2007recurrenceAll}. 

\begin{figure}[ht]
        \centering
        \begin{subfigure}[ht]{\linewidth}
           \begin{subfigure}[h]{0.32\linewidth}
              \includegraphics[width=\linewidth, height=2.5cm]{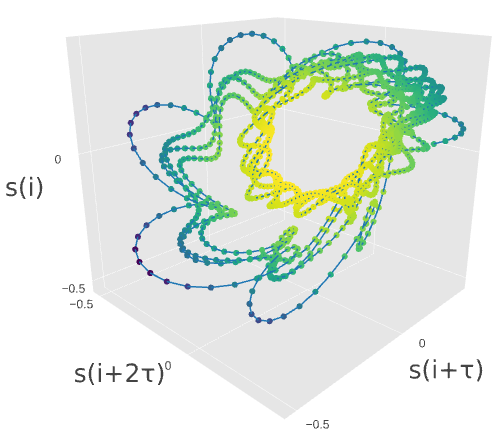}
              \caption{}
              \label{fig:RPS}
          \end{subfigure}
          \begin{subfigure}[h]{0.32\linewidth}
              \includegraphics[width=\linewidth]{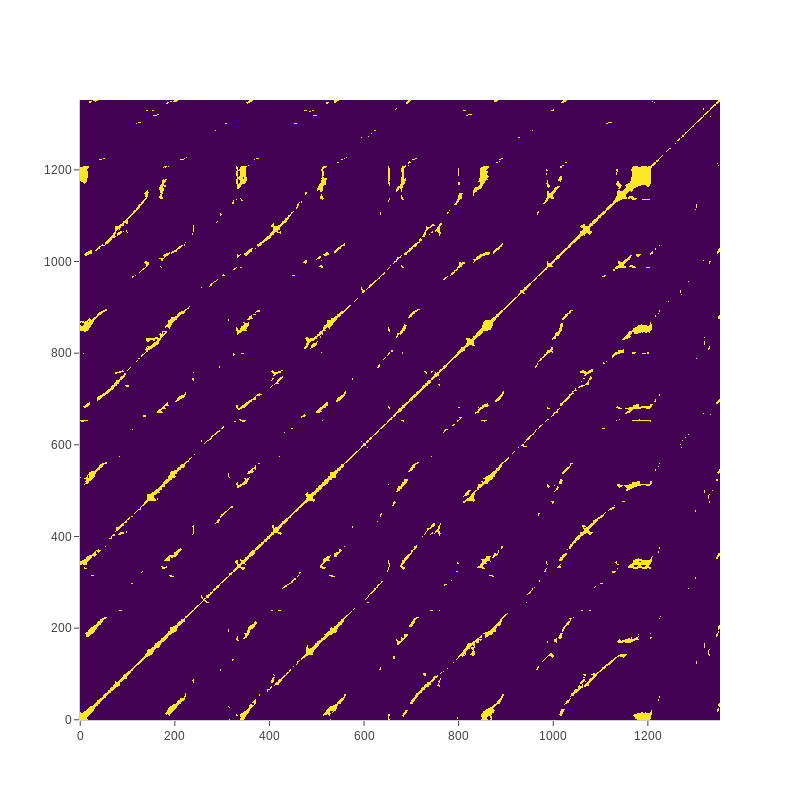}
              \caption{}
              \label{fig:RP}
          \end{subfigure}
          \begin{subfigure}[h]{0.32\linewidth}
              \includegraphics[width=\linewidth]{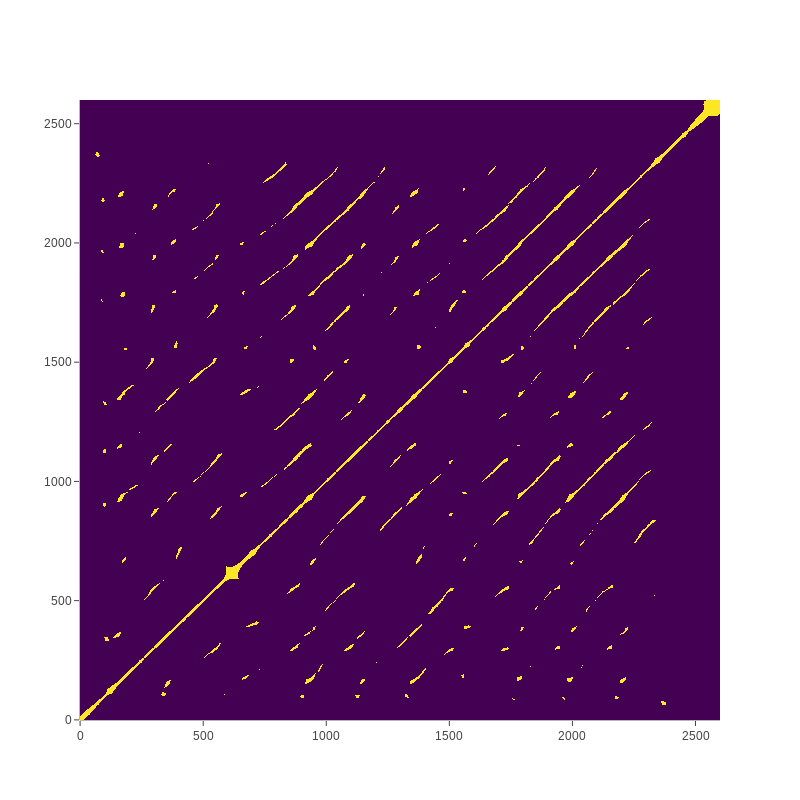}
              \caption{}
              \label{fig:RPLorenz}
          \end{subfigure}
        \end{subfigure}
        \caption{(a) Reconstructed PS ($m=3$, $\tau=7$) and (b) RP ($\epsilon=0.15$, Manhattan norm) of $30$ms frame corresponding to vowel /e/. (c) RP of Lorenz96 system displaying chaotic behavior \cite{Lorenz96}}
        \label{fig:RPS-RPvisualization}
        
\end{figure}

Emerging small-scale structures based on lines of ones or zeros reflect the dynamic behavior of the system. For instance, diagonal lines indicate both similar evolution of states for different parts of PS's orbit and  deterministic chaotic dynamics of the system \cite{marwan2007recurrenceAll}. This is also depicted in Figure \ref{fig:RPS-RPvisualization}. 

% The emerging patterns from the vertical $\prod_{k=1}^{k=l}{\mathbf{R}_{i+k,j}}$, horizontal $\prod_{k=1}^{k=l}{\mathbf{R}_{i,j+k}}$ and diagonal $\prod_{k=1}^{k=l}{\mathbf{R}_{i+k,j+k}}$ lines are firmly related to specific behavior              

\subsubsection{Recurrence Quantification Analysis (RQA)}
\label{sect: RQA measures}
For each $N \times N$ RP we extract $12$ RQA measures using the pyunicorn framework \cite{pyunicorn}. Following the notation established in Section~\ref{sect: Recurrence Plot Computation} we provide an overview of these measures in Table \ref{t:RQAMeasures}; they are comprehensively studied in \cite{marwan2007recurrenceAll}, \cite{RQAExtended2005}. 

\begin{table}[ht]
	\caption{Recurrence Quantification Analysis Measures}
    \label{t:RQAMeasures}
	\centering
    \renewcommand{\arraystretch}{1.6}
    %\footnotesize
    %\fontsize{9}{9}\selectfont

    %\resizebox{\columnwidth}{!} {
    \begin{tabular}{|c|c|}
	    %\specialrule{.1em}{0em}{0em} 
        \hline
        \textbf{Name} & \textbf{Formulation} \\
        \hline
    	Recurrence Rate  & $\frac{1}{N^2}\sum_{i,j = 1}^N \mathbf{R}_{i,j}$ \\
        Determinism & $\frac{\sum_{l=d_{m}}^{N}lP_d(l)}{\sum_{l=1}^{N}lP_d(l)}$ \\
        Max Diagonal Length & 
        $max(\{l_i\}_{i=1}^{N_d})$  \\     
        Average Diagonal Length & 
        $\frac{\sum_{l=d_{m}}^{N}lP_d(l)}{\sum_{l=d_{m}}^{N}P_d(l)}$  \\ 
        Diagonal Entropy & $\sum_{l=d_m}^N \frac{P_d(l)}{N_d}ln(\frac{N_d}{P_d(l)})$ \\
        Laminarity & $\frac{\sum_{l=v_{m}}^{N}lP_v(l)}{\sum_{l=1}^{N}lP_v(l)}$ \\
        Max Vertical Length & 
        $max(\{v_i\}_{i=1}^{N_v})$  \\     
        Trapping Time & 
        $\frac{\sum_{l=v_{m}}^{N}lP_v(l)}{\sum_{l=v_{m}}^{N}P_v(l)}$  \\ 
        Vertical Entropy & $\sum_{l=v_m}^N \frac{P_v(l)}{N_v}ln(\frac{N_v}{P_v(l)})$ \\
        Max White Vertical Length & 
        $max(\{w_i\}_{i=1}^{N_w})$  \\     
        Average White Vertical Length & 
        $\frac{\sum_{l=w_{m}}^{N}lP_w(l)}{\sum_{l=w_{m}}^{N}P_w(l)}$  \\ 
        White Vertical Entropy & $\sum_{l=w_m}^N\frac{P_w(l)}{N_w}ln(\frac{N_w}{P_w(l)})$ \\
        \hline
	\end{tabular}
    %}
\end{table}

\subsubsection{Statistical Functionals}
\label{sect: RQA stats}
After the extraction of frame-wise features and their associated deltas we apply $18$ statistical functionals: min, max, mean, median, variance, skewness, kurtosis, range,  $1_{st}$, $5_{th}$, $25_{th}$, $50_{th}$, $75_{th}$, $95_{th}$, $99_{th}$ percentile and $3$ quartile ranges.     

\subsection{Fused Feature Set (RQA + IS10 Set)}
\label{sect:FusedFeatureSet}
For any emotional speech segment or utterance we extract both feature sets IS10 and RQA as described previously and concatenate them. The final feature vector has $2014$ dimensions. 

\section{Classification Methods}
\label{sect:ClassificationMethods}
%In order to assess the performance of the proposed RQA and IS10 feature sets as well as their fusion, 
We investigate both  utterance-based and  segment-based   SER  as outlined below: \vspace*{1mm} %We also outline the configuration parameters used for the extraction of the RQA feature set. 

\noindent\textbf{Utterance-based method:} For each utterance we obtain its statistical representation by extracting the corresponding feature set as described in Section~\ref{sect:FeatureExtraction}. For emotion classification we employ an SVM with Radial Base Function (RBF) kernel and one-versus-rest LR classifier. Cost coefficient $C$ lies in the interval $[0.001, 30]$ for both SVM and LR models which is the only hyper-parameter to be tuned. Both models are implemented using the scikit-learn framework \cite{scikit-learn}. \vspace*{1mm}

\noindent\textbf{Segment-based method:} We break each utterance into segments of $1.0$ s length and $0.5$ s stride in accordance with \cite{tzinis2017segment}. For each speech segment we extract the feature sets described in Section~\ref{sect:FeatureExtraction} and as a result each utterance is now represented by a sequence of statistical vectors corresponding to different time steps. This sequence is fed as an input to a Long Short Time Memory (LSTM) unit for  emotion classification. SER can be formulated as a many-to-one sequence learning where the expected output of each sequence of segment features is an emotional label derived from the activations of the last hidden layer \cite{shri2016attention}. We employ an A-BLSTM architecture \cite{mirsamadi2017attention} where the decision for the emotional label is derived from a weighted aggregation of all timesteps. We implement this architecture in pytorch \cite{pytorch}. In addition, the grid space of hyper-parameters consists of: number of layers $\{1,2\}$, number of hidden nodes $\{128,256\}$, input noise $[0.3,0.8]$, dropout rate $[0.3,0.8]$ and learning rate $[0.0002, 0.002]$.    

% For utterance-level emotion classification we use SVM with RBF kernel and one versus all Logistic Regression (LR). We perform grid search to select the optimal regularization terms $C$ for SVM and LR and present the best performing models for each experiment.
% For segment-level classification we use an LSTM with 1 layer and a self-attention mechanism \cite{DBLP:journals/corr/BahdanauCB14}. To improve generalization capabilities we also add Dropout of $0.3$ and Gaussian noise in the input with $\sigma=0.2$.

\section{Experiments and Results}

The following databases are used in our experiments: \vspace*{1mm}

\noindent\textbf{SAVEE:} Surrey Audio-Visual Expressed Emotion (SAVEE) Database \cite{SAVEE} is composed of emotional speech voiced by $4$ male actors. SAVEE includes $480$ utterances ($120$ utterances per actor) of $7$ emotions i.e., $60$ anger, $60$ disgust, $60$ fear, $60$ happiness, $60$ sadness, $60$ surprise and $120$ neutral.\vspace*{1mm}

\noindent\textbf{Emo-DB:} Berlin Database of Emotional Speech (Emo-DB) \cite{EmoDB} contains $535$ emotional sentences in German, voiced by $10$ actors ($5$ male and $5$ female). Specifically, $7$ emotions are included i.e., $127$ anger, $45$ disgust, $70$ fear, $71$ joy, $60$ sadness, $81$ boredom and $70$ neutral.\vspace*{1mm}

\noindent\textbf{IEMOCAP:} IEMOCAP database \cite{IEMOCAP} contains $12$ hours of video data of scripted and improvised dialog recorded from $10$ actors. Utterances are organized in $5$ sessions of dyadic interactions between $2$ actors. For our experiments we consider $5531$ utterances including 4 emotions (1103 angry, 1636 happy, 1708 neutral and 1084 sad), where we merge excitement and happiness class into the latter one \cite{emily2017regionalsaliency}, \cite{xia2017multi}, \cite{fayek2017stateoftheart}, \cite{ghosh2016representation}. \vspace*{1mm}

We evaluate our proposed feature set under three different SER tasks described next. We also compare our results with the most relevant experimental setups reported in the literature. For all tasks, we report: Weighted Accuracy (WA) which is the percentage of correct classification decisions and Unweighted Accuracy (UA) which is calculated as the average of recall percentage for each emotional class. 

After an extensive study of the RQA configuration parameters described in Section~\ref{sect: Recurrence Plot Computation}, we conclude that best results on SER tasks are obtained using a frame duration of $20$ ms for extracting RPs. In addition, the best performing parameters for the RP configuration seem to be a Manhattan norm with a threshold setting depending on a fixed recurrence rate lying in $[0.1, 0.2]$.      
% We test a variety of frame durations $\{20, 30, 50\}$ ms from which RPs are computed. For the configuration of the RP we explore the sets of parameters detailed in Section~\ref{sect: Recurrence Plot Computation}. Specifically, we test Manhattan, Euclidean and Supremum norms as well as multiple selection criteria for the threshold value $\epsilon$, depending on ad-hoc threshold setting, fixed recurrence rate and fixed $\sigma$. The respective ratio parameter for the three aforementioned criteria lies in $[0.05,0.5]$. We report that the best configurations for the RQA set is      

\subsection{Speaker Dependent (SD)}
\label{sect:SDExperiments}
We evaluate RQA features on SAVEE and Emo-DB following the utterance-based approach described in Section~\ref{sect:ClassificationMethods}. In this setup we apply per-speaker $z$-normalization (PS-N) and split randomly utterances in train and test sets. Accuracies using $5$-fold cross-validation are summarized on Table~\ref{t:sd} for the best performing classifier hyper-parameter values.

The fused set achieves significant performance improvement over the baseline IS10 feature set for both datasets. On SAVEE, WA is improved by $3.1\%$ ($77.1\% \rightarrow 80.2\%$)  and UA by $3.4\%$ ($74.5\% \rightarrow 77.9\%$).  We also achieve an improvement of $4.9\%$ ($88.4\% \rightarrow 93.3\%$) and $5.7\%$ ($87.2\% \rightarrow 92.9\%$) for WA and UA, respectively on Emo-DB. The feature set used in \cite{sun2017ensemble} is extracted over cepstral, spectral and prosodic LLDs similar to the ones used in IS10 \cite{schuller2010interspeech}. Noticeably, they achieve similar performance to ours when we use only IS10 but our fused set with LR outperforms on both Emo-DB ($5\%$ in UA and $4.6\%$ in WA) and SAVEE ($4.5\%$ in UA and $3.9\%$ in WA). The proposed combination of features and LR also surpasses a Convolutional SAE approach \cite{mao2014autoencoder} in terms of WA by $5\%$ on Emo-DB and $4.8\%$ on SAVEE. Presumably, RQA measures contain information closely related to speaker-specific emotional dynamics not captured by conventional features.

\begin{table}[h]
	\caption{SD results on SAVEE and Emo-DB. (ESR) Ensemble Softmax Regression}
    \label{t:sd}
	\centering
    %\footnotesize
    %\fontsize{9}{9}\selectfont

    %\resizebox{\columnwidth}{!} {
    \begin{tabular}{cccccc}
	    %\specialrule{.1em}{0em}{0em} 
        \hline
    	\multirow{2}{*}{Features} & \multirow{2}{*}{Model}  & \multicolumn{2}{c}{SAVEE} & \multicolumn{2}{c}{Emo-DB} \\
        %\specialrule{.1em}{0em}{0em}
                                  &     & WA  & UA   & WA & UA  \\ 
		\hline
        %\specialrule{.1em}{0em}{0em}
        \multirow{2}{*}{IS10}     & SVM & 77.1 & 74.5 & 88.4 & 87.2\\ 
                                  & LR  & 74.4 & 71.8 & 87.4 & 86.3 \\ 
        %\specialrule{.1em}{0em}{0em}
        \hline
		\multirow{2}{*}{RQA}      & SVM & 66.0 & 63.0 & 81.8 & 80.4\\ 
                                  & LR  & 64.4 & 61.1 & 81.9 & 79.9 \\ 
        %\specialrule{.1em}{0em}{0em}
		\hline
        \multirow{2}{*}{\shortstack{RQA+IS10}} & SVM & 77.3 & 75.5 & 90.1 & 88.9 \\ 
                                  & LR  & \textbf{80.2} & \textbf{77.9} & \textbf{93.3} & \textbf{92.9} \\
	    %\specialrule{.1em}{0em}{0em} 
        \hline
        \cite{mao2014autoencoder} Spectrogram & SAE & 75.4 & {\raggedleft-}  & 88.3 & {\raggedleft-}   \\
        \cite{sun2017ensemble} LLDs Stats &  ESR & 76.3 & 73.4 & 88.7 & 87.9 \\	
        
        \hline
        
	\end{tabular}
    %}
\end{table}

% \begin{table}[h]
% 	\caption{Speaker dependent classification results of WA and UA}
%     \label{t:sd}
% 	\centering
%     %\footnotesize
%     %\fontsize{9}{9}\selectfont

%     %\resizebox{\columnwidth}{!} {
%     \begin{tabular}{cccccc}
% 	    %\specialrule{.1em}{0em}{0em} 
%         \hline
%     	\multirow{2}{*}{Features} & \multirow{2}{*}{Model}  & \multicolumn{2}{c}{SAVEE} & \multicolumn{2}{c}{Emo-DB} \\
%         %\specialrule{.1em}{0em}{0em}
%                                   &     & WA  & UA   & WA & UA  \\ 
% 		\hline
%         %\specialrule{.1em}{0em}{0em}
%         \multirow{2}{*}{IS10}     & SVM & 77.08 & 74.52 & 88.36 & 87.22\\ 
%                                   & LR  & 74.37 & 71.75 & 87.40 & 86.33 \\ 
%         %\specialrule{.1em}{0em}{0em}
%         \hline
% 		\multirow{2}{*}{RQA}      & SVM & 66.04 & 62.98 & 81.84 & 80.35\\ 
%                                   & LR  & 64.38 & 61.07 & 81.86 & 79.94 \\ 
%         %\specialrule{.1em}{0em}{0em}
% 		\hline
%         \multirow{2}{*}{\shortstack{RQA+IS10}} & SVM & 77.29 & 75.48 & 90.10 & 88.93 \\ 
%                                   & LR  & \textbf{80.21} & \textbf{77.86} & \textbf{93.27} & \textbf{92.87} \\
% 	    %\specialrule{.1em}{0em}{0em} 
%         \hline
% 	\end{tabular}
%     %}
% \end{table}

\subsection{Speaker Independent (SI)}
\label{sect:SIExperiments}
Again, we follow the utterance-based approach described in Section~\ref{sect:ClassificationMethods} on both SAVEE and Emo-DB datasets but we do not make any assumptions for the identity of the user during training. We use leave-one-speaker-out cross validation, where one speaker is kept for testing and the rest for training. The mean and standard deviation are calculated only on training data and used for $z$-normalization on all data. From now on we refer to this normalization as Per Fold-Normalization (PF-N). Table~\ref{t:si} presents accuracies averaged over all folds for the best performing classifier hyper-parameter values.

In comparison with the baseline IS10 feature set, the fused feature set obtains an absolute improvement of $5.5\%$ and $8.2\%$ on SAVEE as well as $2.4\%$ and $3.2\%$ on Emo-DB in terms of WA and UA, respectively. Furthermore, our fused set achieves higher performance on SAVEE ($3.5\%$ in WA and $4.5\%$ in UA) and slightly lower in Emo-DB compared to \cite{sun2017ensemble}. In \cite{sun2015weightedHuMoments} Weighted Spectral Features based on Hu Moments (WSFHM) are fused with IS10 on utterance-level which is similar to our approach. In direct comparison using the same model (SVM) we surpass the reported performance in terms of WA by $2.5\%$ and $0.4\%$ on SAVEE and Emo-DB, respectively. In addition, both RQA and IS10 sets achieve quite low performance on SAVEE. However, their combination yields an impressive performance improvement of $5.5\%$ ($48.5\% \rightarrow 54.0\%$) in WA and $10.7\%$ ($43.1\% \rightarrow 53.8\%$) in UA over IS10 when we use LR. Our results suggest that RQA measures preserve invariant aspects of nonlinear dynamics occurring in emotional speech and are shared across different speakers.        

\begin{table}[h]
	\caption{SI results on SAVEE and Emo-DB. (ESR) Ensemble Softmax Regression}
    \label{t:si}
	\centering
    %\footnotesize
    %\fontsize{9}{9}\selectfont

    %\resizebox{\columnwidth}{!} {
    \begin{tabular}{cccccc}
	    %\specialrule{.1em}{0em}{0em} 
        \hline
    	\multirow{2}{*}{Features} & \multirow{2}{*}{Model}  & \multicolumn{2}{c}{SAVEE} & \multicolumn{2}{c}{Emo-DB} \\
        %\specialrule{.1em}{0em}{0em}
                                  &     & WA & UA   & WA & UA \\ 
        %\specialrule{.1em}{0em}{0em}
        \hline
        \multirow{2}{*}{IS10}     & SVM & 47.5 & 45.6 & 79.7 & 74.3\\ 
                                  & LR  & 48.5 & 43.1 & 76.1 & 71.9 \\ 
        %\specialrule{.1em}{0em}{0em}
        \hline
		\multirow{2}{*}{RQA}      & SVM & 45.6 & 41.1 & 70.9 & 64.2\\ 
                                  & LR  & 47.7 & 42.3 & 71.1 & 67.1 \\ 
        %\specialrule{.1em}{0em}{0em}
		\hline
        \multirow{2}{*}{\shortstack{RQA+IS10}} & SVM & 52.5 & 50.6 & 82.1 & 76.9 \\ 
                                               & LR  & \textbf{54.0} & \textbf{53.8} & 80.1 & 77.5 \\
	    %\specialrule{.1em}{0em}{0em} 
        \hline
	    \cite{sun2017ensemble} LLDs Stats &  ESR &
	51.5 & 49.3 & \textbf{82.4} & \textbf{78.7} \\
    \cite{sun2015weightedHuMoments} WSFHM+IS10 & SVM & 50.0 & {\raggedleft-} & 81.7 &  {\raggedleft-} \\
    \hline
	\end{tabular}
    %}
\end{table}

% \begin{table}[h]
% 	\caption{Speaker independent results}
%     \label{t:si}
% 	\centering
%     %\footnotesize
%     %\fontsize{9}{9}\selectfont

%     %\resizebox{\columnwidth}{!} {
%     \begin{tabular}{cccccc}
% 	    %\specialrule{.1em}{0em}{0em} 
%         \hline
%     	\multirow{2}{*}{Features} & \multirow{2}{*}{Model}  & \multicolumn{2}{c}{SAVEE} & \multicolumn{2}{c}{Emo-DB} \\
%         %\specialrule{.1em}{0em}{0em}
%                                   &     & WA & UA   & WA & UA \\ 
%         %\specialrule{.1em}{0em}{0em}
%         \hline
%         \multirow{2}{*}{IS10}     & SVM & 47.50 & 45.59 & 79.67 & 74.34\\ 
%                                   & LR  & 48.54 & 43.09 & 76.11 & 71.92 \\ 
%         %\specialrule{.1em}{0em}{0em}
%         \hline
% 		\multirow{2}{*}{RQA}      & SVM & 45.62 & 41.07 & 70.90 & 64.24\\ 
%                                   & LR  & 47.71 & 42.26 & 71.05 & 67.05 \\ 
%         %\specialrule{.1em}{0em}{0em}
% 		\hline
%         \multirow{2}{*}{\shortstack{RQA+IS10}} & SVM & 52.50 & 50.63 & \textbf{82.08} & 76.88 \\ 
%                                                & LR  & \textbf{53.96} & \textbf{53.75} & 80.05 & \textbf{77.45} \\
% 	    %\specialrule{.1em}{0em}{0em} 
%         \hline
% 	\end{tabular}
%     %}
% \end{table}

\subsection{Leave One Session Out (LOSO)}
In this task, we assume that the test-speaker identity is unknown but we are able to train our model considering other speakers who are recorded in similar conditions. We evaluate on both utterance and segment-based methods (described in Section~\ref{sect:ClassificationMethods}) on IEMOCAP. Given our assumption, we treat each of the $5$ sessions as a speaker group \cite{IEMOCAP}. We use LOSO in order to create train and test folds. In each fold, we use $4$ sessions for training and the remaining $1$ for testing. For the testing session we use one speaker as testing set and the other for tuning the hyper-parameters of our models. We repeat the evaluation by reversing the roles of the two speakers. 
In the final assessment, we report the average performance obtained in terms of WA and UA obtained from all speakers \cite{emily2017regionalsaliency}, \cite{xia2017multi}, \cite{ghosh2016representation}. In order to be easily comparable with the literature we follow three different normalization schemes. We use the aforementioned PS-N and PF-N schemes as well as Global $z$-normalization (G-N). In G-N we calculate the global mean and standard deviation from all the available samples in the dataset and perform $z$-normalization over them. Results on IEMOCAP for the three different normalization schemes are demonstrated on Table \ref{t:iemo}.

A consistent performance improvement is shown for all combinations of normalization techniques and employed models when the fused set is used instead of IS10. Specifically, for SVM the fused set yields a relative improvement varying from $0.3\%$ to $1.0\%$ in WA and from $0.2\%$ to $0.9\%$ in UA under all normalization strategies. The same applies for LR (in WA from $0.8\%$ to $1.0\%$ and in UA from $0.3\%$ to $1.0\%$ ) as well as for A-BLSTM (in WA from $0.1\%$ to $0.7\%$ and in UA from $0.2\%$ to $0.7\%$). In accordance with our intuition \cite{tzinis2017segment}, a segment-based approach using A-BLSTM surpasses all utterance-based ones in WA from $3.4\%$ to $8.4\%$ and in UA from $3.8\%$ to $6.8\%$ for all normalization schemes, when the fused set is used. 

In \cite{emily2017regionalsaliency} low level Mel Filterbank (MFB) features are fed directly to a CNN. In \cite{ghosh2016representation} a stacked autoencoder is used to extract feature representations from spectrograms of glottal flow signals and then a BLSTM is used for classification. We surpass both reported results by $0.2\%$ in UA for \cite{emily2017regionalsaliency} and by a margin of $8.7\%$ in WA and $8.5\%$ in UA for \cite{ghosh2016representation}, respectively even with simple models. 
Compared to a multi-task DBN trained for both discrete emotion classification and for valence-activation in \cite{xia2017multi}, we report $2.0\%$ higher WA and $3.1\%$ higher UA. 
We also report $4.6\%$ higher UA and $1.9\%$ lower WA compared to CNNs over spectrograms \cite{fayek2017stateoftheart}. We assume that this inconsistency in performance metrics occurs because a slightly different experimental setup is followed where the final session is excluded from testing \cite{fayek2017stateoftheart}.

% What the fuck :|
% For IEMOCAP we perform segment-level classification using 2 models, LSTM and LSTM with self-attention (A-BLSTM).
% In order for our results to be comparable with the literature we use LOGSO (or Leave One Session Out) split. Specifically, for each fold, utterances in one session $S$ are kept out from the training set. The utterances of one speaker in $S$ are used for validation while the other speaker is used for testing. This yields $10$  folds and we report the mean UA and WA across folds. We evaluate this split with $3$ normalization schemes found in the literature. The first is to apply per-speaker $z$-normalization (PS-N), described in Section~\ref{sect:SDExperiments}. The second is per-fold $z$-normalization (PF-N), where we normalize all utterances using the mean and standard deviation of samples in the training set. The third normalization scheme is cross-dataset $z$-normalization, where the mean and standard deviation of all utterances are used for normalization. Results are presented in Table~\ref{t:iemo}.

\begin{table}[h]
	\caption{LOSO results on IEMOCAP. (GFS): Glottal Flow Spectrogram, (SP): Spectrogram.}
    \label{t:iemo}
	\centering
    %\footnotesize
    %\fontsize{9}{9}\selectfont

    %\resizebox{\columnwidth}{!} {
    \begin{tabular}{ccK{0.33cm}K{0.33cm}K{0.33cm}K{0.33cm}K{0.33cm}K{0.33cm}}

%     \begin{tabular}{ccP{0.34cm}p{0.34cm}p{0.34cm}p{0.34cm}p{0.34cm}p{0.34cm}}
	    %\specialrule{.1em}{0em}{0em} 
        \hline
    	\multirow{2}{*}{Features} & 
        \multirow{2}{*}{Model}  & 
        \multicolumn{2}{c}{PS-N} & 
        \multicolumn{2}{c}{PF-N} &
        \multicolumn{2}{c}{G-N} \\
		&     & WA & UA & WA & UA & WA & UA\\ 
        \hline
        \multirow{3}{*}{IS10}     
        & SVM & 58.3 & 60.9 & 58.9 & 60.1 & 59.2 & 60.5 \\
        & LR & 57.5 & 61.2 & 54.6 & 57.9 & 53.5 & 57.5 \\
        & A-BLSTM & 62.0 & 65.1 & 62.6 & 65.0 & 62.8 & 65.0 \\
        \hline
		\multirow{3}{*}{RQA}      
        & SVM & 52.9 & 54.6 & 53.1 & 53.8 & 53.1 & 53.7\\
        & LR & 52.2 & 54.8 & 52.6 & 54.0 & 52.8 & 54.3 \\
		& A-BLSTM  & 55.6 & 59.3 & 56.6 & 58.3 & 56.7 & 58.7 \\
		\hline
%         \multirow{3}{*}{RQA + IS10} 
		RQA
        & SVM & 59.3 & 61.8 & 59.2 & 60.4 & 59.5 & 60.7 \\
         +& LR & 58.3 & 62.0 & 55.6 & 58.7 & 54.5 & 58.7 \\
        IS10& A-BLSTM & \textbf{62.7} & \textbf{65.8} & \textbf{63.0} & \textbf{65.2} & 62.9 & \textbf{65.5} \\
        
        \hline 
        \cite{emily2017regionalsaliency} MFB & CNN
& {\raggedleft-} & 61.8 & {\raggedleft-} & {\raggedleft-} & {\raggedleft-} & {\raggedleft-} \\
		%\hline
		\cite{xia2017multi} IS10 & DBN & {\raggedleft-} & {\raggedleft-} & {\raggedleft-} & {\raggedleft-} & 60.9 & 62.4 \\
		%\hline
         \cite{fayek2017stateoftheart} SP & CNN & {\raggedleft-} & {\raggedleft-} & {\raggedleft-} & {\raggedleft-} & \textbf{64.8} & 60.9 \\
        \cite{ghosh2016representation} GFS & BLSTM & {\raggedleft-} & {\raggedleft-} & 50.5 & 51.9 & {\raggedleft-} & {\raggedleft-} \\

        \hline

	\end{tabular}
    %}
\end{table}

\section{Conclusions}
We investigated the usage of nonlinear RQA measures extracted from RPs for SER. The effectiveness of these features has been tested under both utterance-based and segment-based approaches across three emotion databases. The fusion of nonlinear and conventional feature sets yields significant performance improvement over traditional feature sets for all SER tasks; the performance improvement is especially large when speaker identity is unknown. The fused data set improves on the state-of-the-art for SER under most testing conditions, classification methods and datasets. Recurrence analysis of speech signals is a promising direction for SER research. In the future, we plan to automatically extract features from RPs using convolutional autoencoders in order to substitute RQA measures.

\section{Acknowledgements}
This work has been partially supported by the BabyRobot project supported by EU H2020 (grant \#687831). Special thanks to Nikolaos Athanasiou and Nikolaos Ellinas for their contributions on the experimental environment setup.

\bibliographystyle{IEEEtran}

\bibliography{mybib}

% \begin{thebibliography}{9}
% \bibitem[1]{Davis80-COP}
%   S.\ B.\ Davis and P.\ Mermelstein,
%   ``Comparison of parametric representation for monosyllabic word recognition in continuously spoken sentences,''
%   \textit{IEEE Transactions on Acoustics, Speech and Signal Processing}, vol.~28, no.~4, pp.~357--366, 1980.
% \bibitem[2]{Rabiner89-ATO}
%   L.\ R.\ Rabiner,
%   ``A tutorial on hidden Markov models and selected applications in speech recognition,''
%   \textit{Proceedings of the IEEE}, vol.~77, no.~2, pp.~257-286, 1989.
% \bibitem[3]{Hastie09-TEO}
%   T.\ Hastie, R.\ Tibshirani, and J.\ Friedman,
%   \textit{The Elements of Statistical Learning -- Data Mining, Inference, and Prediction}.
%   New York: Springer, 2009.
% \bibitem[4]{YourName17-XXX}
%   F.\ Lastname1, F.\ Lastname2, and F.\ Lastname3,
%   ``Title of your INTERSPEECH 2018 publication,''
%   in \textit{Interspeech 2018 -- 19\textsuperscript{th} Annual Conference of the International Speech Communication Association, September 2-6, Hyderabad, India Proceedings, Proceedings}, 2018, pp.~100--104.
% \end{thebibliography}

\end{document}